\documentclass[twocolumn,subfig]{aastex631}

\usepackage{colortbl}
\usepackage{tabularx}
\usepackage{placeins}
\usepackage{float}
\usepackage{lipsum}
\usepackage{xcolor}
\usepackage{lineno}

\begin{document}

\title[Radio emission of Ser~X--1]{Variable radio emission of neutron star X-ray binary Ser~X--1 during its persistent soft state}

\author{Eli C. Pattie}
\affiliation{Department of Physics and Astronomy, Texas Tech University, Lubbock, TX 79409-1051, USA}
\author{Thomas J. Maccarone}
\affiliation{Department of Physics and Astronomy, Texas Tech University, Lubbock, TX 79409-1051, USA}
\author[0000-0003-3906-4354]{Alexandra J. Tetarenko}
\affiliation{Department of Physics and Astronomy, University of Lethbridge, Lethbridge, Alberta, T1K 3M4, Canada}
\author[0000-0003-3124-2814]{James C.A.  Miller-Jones}
\affiliation{International Centre for Radio Astronomy Research, Curtin University, GPO Box U1987, Perth, WA 6845, Australia}
\author{Manuel Pichardo Marcano}
\affiliation{Department of Physics and Astronomy, Vanderbilt University, 6301 Stevenson Center Lane, Nashville, TN 37235, USA}
\affiliation{Department of Life and Physical Sciences, Fisk University, 1000 17th Avenue N., Nashville, TN 37208, USA}
\author{Liliana E. Rivera Sandoval}
\affiliation{Department of Physics and Astronomy, University of Texas Rio Grande Valley, Brownsville, TX 78520, USA}



\begin{abstract}

Ser~X--1 is a low mass neutron star X-ray binary and has been persistently accreting since its discovery in the 1960s. It has always been observed to be in a soft spectral state and has never showed substantial long-term X-ray variability. Ser~X--1 has one previous radio observation in the literature in which radio emission was detected during this soft state, which is contrary to the behavior of black hole X-ray binaries. We have recently obtained 10 randomly sampled radio epochs of Ser~X--1 in order to further investigate its anomalous soft state radio emission. Out of 10 epochs, we find 8 non-detections and 2 detections at 10~GHz flux densities of $19.9\pm4.2$~$\mu$Jy and $32.2\pm3.6$~$\mu$Jy. We do not detect polarization in either epoch, ruling out very high polarization levels ($\lesssim$~63\% and 34\%). We compare these Ser~X--1 results to other X-ray binaries and consider explanations for its long term variable radio behavior.

\end{abstract}



\section{Introduction}

X-ray binaries (XRBs) are systems where a neutron star or black hole accretes matter from a companion star. Most XRBs have low mass companion stars (LMXBs) where accretion occurs via Roche lobe overflow, in which case the accreted material flows into an accretion disk around the compact object (except intrinsically high magnetic field neutron stars, where the magnetic field may prevent the formation of a disk), propagating inward toward the accretor over time. Material from the inner accretion disk may then either be accreted onto the compact object or be accelerated away from the binary in the form of winds and jets \citep{Fender2004,Remillard2006,Done2007}. In general, black hole and (weak magnetic field) neutron star X-ray binaries (BHXBs; NSXBs) with low mass companions are similar to each other in terms of their geometries and their accretion behaviors \citep{Narayan1995}. Notably, they make spectral state transitions at similar fractions of the Eddington luminosity ($\mathrm{L_{Edd}}$) \citep{Maccarone2003}. However, they usually differ significantly in their radio luminosities, with NSXBs generally being fainter by about two orders of magnitude than BHXBs for similar X-ray luminosities \citep{Migliari2006,vandenEijnden2021}. Radio emission is dominated by the non-thermal jets, indicating that neutron stars are less efficient at producing jets than black holes. It is well-established from observations that radio luminosity is closely correlated to X-ray luminosity, i.e., jet power depends on the mass accretion rate, as expected for the accretion flow feeding material to the jet \citep[e.g.,][]{Hannikainen1998,vandenEijnden2018}.

\begin{table*}
\centering
    \setlength{\tabcolsep}{7pt}
	\begin{tabular}{cccccccc}
	    \hline
	    Project Code & Date & MJD & Config. & On-source & Obs. & Detection peak  & RMS \\
         & (YYYY MM DD) & (Obs. start) &  & (minutes) & \# & ($\mu$Jy/beam) & ($\mu$Jy/beam) \\
		\hline\hline
		21A-175 & 2021 06 09 & 59374.3 & C & 44 & 1 & $< 14.1$ & 4.7 \\
        \arrayrulecolor{lightgray}\hline
        22B-202 & 2022 10 08 & 59860.9 & C & 44 & 2 & $< 13.2$ & 4.4 \\
        22B-202 & 2022 11 04 & 59887.8 & C & 44 & 3 & $< 11.7$ & 3.9 \\
        22B-202 & 2022 11 27 & 59910.8 & C & 44 & 4 & {\bf 19.9 $\pm$ 4.2} & 4.2 \\
        \hline
        23A-343 & 2023 04 19 & 60053.4 & B & 44 & 5 & $< 11.4$ & 3.8 \\
        23A-343 & 2023 04 29 & 60063.4 & B & 44 & 6 & $< 10.5$ & 3.5 \\
        23A-343 & 2023 05 08 & 60072.3 & B & 44 & 7 & $< 10.5$ & 3.5 \\
        \hline
        23A-343 & 2023 05 15 & 60079.3 & B & 44 & 8 & $< 10.8$ & 3.6 \\
        23A-343 & 2023 05 26 & 60090.3 & B & 44 & 9 & {\bf 32.2 $\pm$ 3.6} & 3.6 \\
        23A-343 & 2023 06 02 & 60097.3 & BnA & 44 & 10 & $< 10.5$ & 3.5 \\        
		\arrayrulecolor{black}\hline		
        All non-dets &  &  &  & 352 &  & $< 4.5$ &  1.5 \\
        \hline	
        *AR476 & 2002 05 27 & 52421.6 & A & 221  & & 68.7 $\pm$ 21 & 21 \\
        \hline	
	\end{tabular}
	\caption{Observations and results from imaging of Ser~X--1. Out of 10 total epochs, we find 2 detections. 
 The upper limits for non-detections are given at the 3$\sigma$ value ($3~\cdot~$RMS from same row), and errors on the detections are $1~\cdot$~RMS. 
 *AR476 are archival data (classic VLA), presented in \citet{Migliari2004}, which we have re-reduced independently for comparison with our recent data.}
 
	\label{table:observations}
\end{table*}

Jets are nearly ubiquitously present during ``hard'' accretion states, at relatively low accretion rates ($\lesssim0.01\mathrm{L_{Edd}})$ when the inner disk is geometrically thick and is canonically truncated a distance away from the accretor (e.g., \citet{Esin2001,Zdziarski2004,Zdziarski2022}, though also see, e.g., \citet{Miller2006}). 
During ``soft'' accretion states at higher accretion rates 
($\sim0.1-<\mathrm{L_{Edd}}$) with a geometrically thin disk that extends close to the accretor, radio emission is generally not detected among XRBs, with only a few stray detections in the literature (aside from the unique case of 4U~1820--30, further described below).
This indicates that jets are strongly suppressed by up to multiple orders of magnitude during these times (e.g., at least 3.5 orders of magnitude in a BHXB from \citealt{Russell2019}, and very deep upper limits from \citet{Maccarone2020}), with some suggestions of complete physical disappearance. There is also a third regime at very high accretion rates ($\gtrsim\mathrm{L_{Edd}}$) where radio jets are commonly observed (e.g., the ``Z-source" type of NSXB \citep{Migliari2006}). However, these systems inhabit a different type of soft state where the geometry of the inner disk is likely significantly affected by radiation pressure at these very high accretion rates, and is not comparable to the canonical soft state that is of interest here.

However, there are two persistently accreting NSXBs that have had radio emission detected during a canonical soft state. These are 4U~1820--30 and Ser~X--1, both of which are low magnetic field neutron stars with low mass companions. Both systems' soft state radio emission was initially presented in \citet{Migliari2004}. Since then, 4U~1820--30 has had numerous additional radio observations, and in every epoch there has indeed been radio emission detected at $\gtrsim 30~\mu$Jy, indicating that this source is persistently producing radio emission even during soft states (e.g., \citealt{Russell2021}), though the radio emission mechanism has not been unambiguously identified. In contrast, Ser~X--1 has no subsequent radio observations in the literature. Motivated by the sole previous VLA radio observation and detection of Ser~X--1 in a soft state two decades ago, and by the multiple and reliable radio detections of 4U~1820--30 in its soft state in the years since, we have carried out a radio campaign to significantly increase the number of radio observations of Ser~X--1 to obtain a more complete idea of its radio emission behavior over time, e.g., whether it is persistent like 4U~1820--30, or if it is intermittent or strongly variable.

\begin{figure*}[ht]
    \centering
    \includegraphics[width=\textwidth]{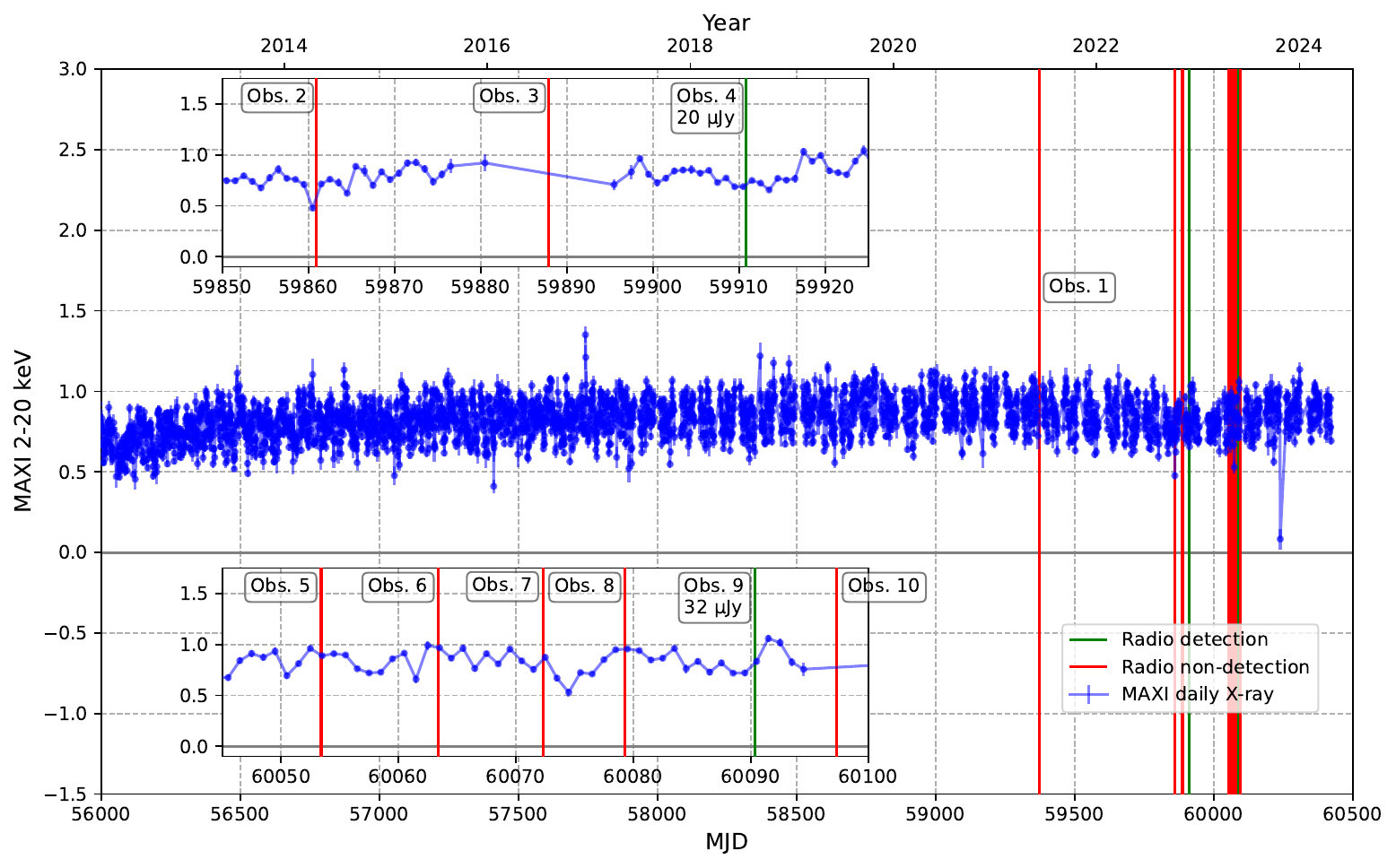}\\
    \caption{All daily MAXI data for Ser~X--1 (after removing bad data points, defined as the error value being more than 3 standard deviations from the median error value). The MAXI data demonstrate the very consistent X-ray behavior of Ser~X--1 over the past $\sim$10 years. Vertical lines indicate times of the 10 recent VLA observations: red lines are non-detections and green are detections. The two insets zoom in on the two groups of multiple observations. Obs. 1 occurred during a several-days-long gap in MAXI data (as did Obs. 3 and Obs. 10), so we do not show a zoomed-in segment for Obs. 1.}
    \label{fig:MAXIall_inset}
\end{figure*}

Ser~X--1 is a low-mass neutron star X-ray binary that is persistently accreting material from its companion, likely in a 2~hour orbit with an M dwarf companion and observed at a very low inclination angle \citep{Cornelisse2013}, at a distance of $\sim$7.7~kpc \citep{Galloway2008}. Its X-ray flux has been very stable since its discovery in the 1960s \citep{Friedman1967}, is always observed in a soft accretion state \citep[e.g.,][]{Matranga2017,Ludlam2018,Mondal2020}, and occasionally exhibits thermonuclear X-ray bursts, including some superbursts \citep{Li1977,Cornelisse2002}. Its X-ray emission has never shown strong variability, especially as seen over the last decade with the Monitor of All-sky X-ray Image (MAXI; \citealt{Matsuoka2009}, Figure~\ref{fig:MAXIall_inset}), where its X-ray emission has been remarkably constant. However, we find 2 radio detections over 10 total epochs over two years with no apparent correlation with X-ray luminosity, a scenario that is difficult to reconcile with XRBs.

\section{Observations and Data Reduction}

Ser~X--1 was observed over 10 epochs from 2021 to 2023 with the Very Large Array in X-band (8--12~GHz) under Project Codes 21A--175, 22B-202, and 23A-343. All 10 observations had identical scheduling blocks (i.e. equal on-source times), and with a polarization calibration setup. The flux and polarization angle calibrator was 3C286 (J1331+3030), the polarization leakage calibrator was J1407+2827, and the phase calibrator was J1824+1044. Observations are listed in Table~\ref{table:observations}. The data for each observation were obtained from the archive with NRAO's CASA pipeline calibration applied (Science Ready Data Products). The target data were further flagged in CASA with \texttt{rflag} and \texttt{tfcrop} and inspected visually. A Stokes~I image was manually cleaned with \texttt{tclean} per observation with \texttt{weighting=natural} and \texttt{pblimit=0.1} to include a nearby bright source in the image and suppress its artifacts through deconvolution. Observations with radio detections were cleaned a second time with \texttt{weighting=briggs} and \texttt{briggs=0.5} to more clearly confirm the shape of the detected source and further suppress artifacts, and a third time with full Stokes polarization after performing the standard polarization calibration process to the measurement set. Another image was cleaned combining all of the non-detection observations as well. Positions of the the two detections and their \texttt{uvmodelfit} flux densities are provided in Table~\ref{table:fits}, and a list of other sources in the field of view is provided in Table~\ref{table:otherSources}.

\begin{figure*}[ht]
\centering
\begin{tabular}{cc}
{\includegraphics[width=0.43\textwidth]{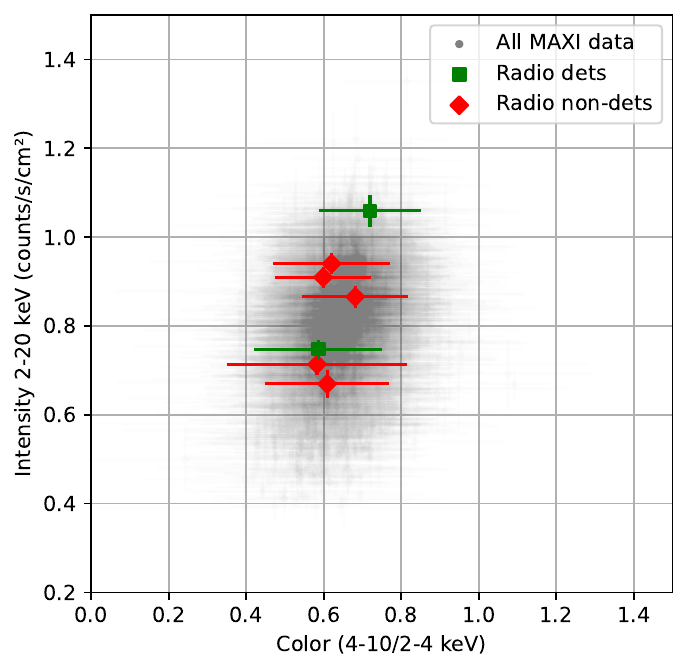}} 
{\includegraphics[width=0.45\textwidth]{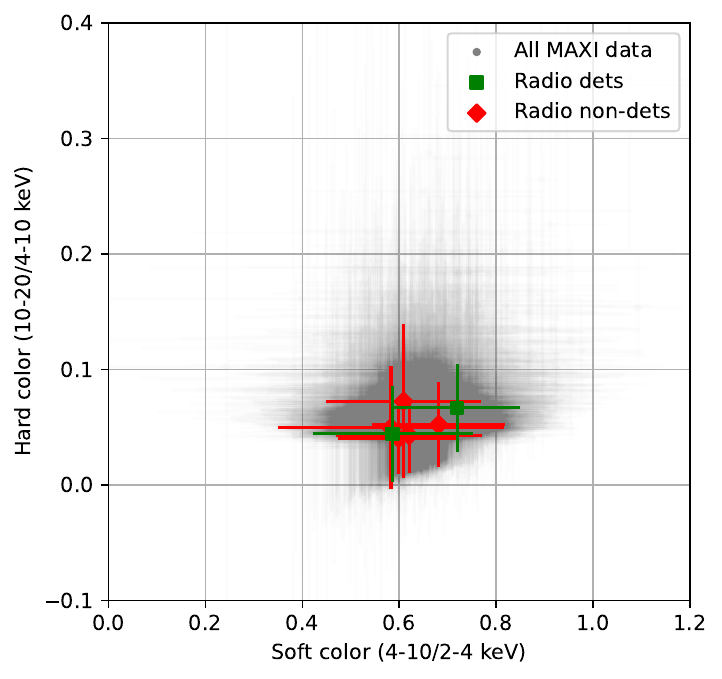}} \\
\end{tabular}
\caption{Hardness-intensity (HID, {\it left}) and color-color (CCD, {\it right}) diagrams of Ser~X--1 from MAXI data (energy bands of 2--4, 4--10, and 10--20 keV; bad data points were removed similar to Figure~\ref{fig:intra}) All daily MAXI data points are plotted in grey circles (transparency alpha value of 0.01 to more clearly show density of data points). MJD days with radio detections are plotted in green squares, and radio non-detections in red diamonds. Three of the eight non-detection epochs are missing MAXI data coverage, as noted in Figure~\ref{fig:MAXIall_inset}. In both diagrams, we do not find that the radio detections occur preferentially in any particular region, relative both to all data points and the other radio non-detection epochs.}
\label{fig:tracks}
\end{figure*}

\begin{figure*}[ht]
\centering
\begin{tabular}{cc}
{\includegraphics[width=0.45\textwidth]{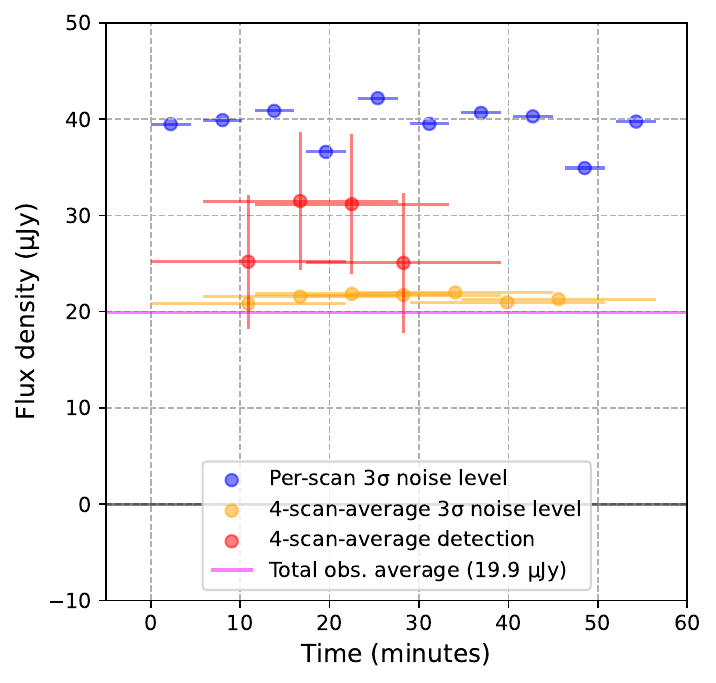}} 
{\includegraphics[width=0.45\textwidth]{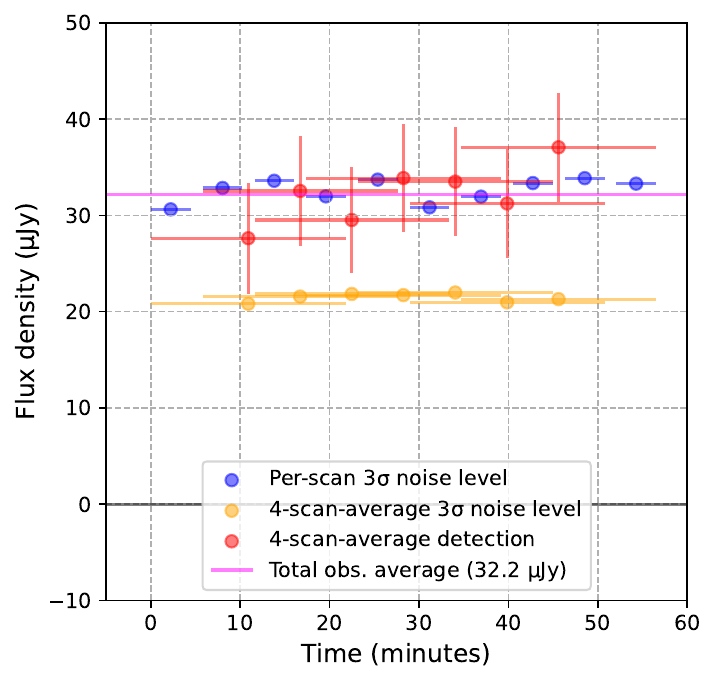}} \\
\end{tabular}
\caption{Intra-observation analyses of the two Ser~X--1 detections. Blue bars represent the per-scan 3$\sigma$ noise limits; we do not clearly detect Ser~X--1 in any individual scan. Orange bars are 3$\sigma$ noise levels of a rolling average of 4 scans, shown for all 4-scan sets. Red markers are 4-scan-average detections. Horizontal error bars represent the time integrated over for the corresponding data. The magenta horizontal line in each plot is the nominal total epoch detection flux density value for reference. We do not find strong variability occurring over $\sim$1 hour in each detection epoch based on these results, though there may be some slow drifting.}
\label{fig:intra}
\end{figure*}

\begin{figure*}[ht]
\centering
\begin{tabular}{ccc}
{\includegraphics[width=0.3\textwidth]{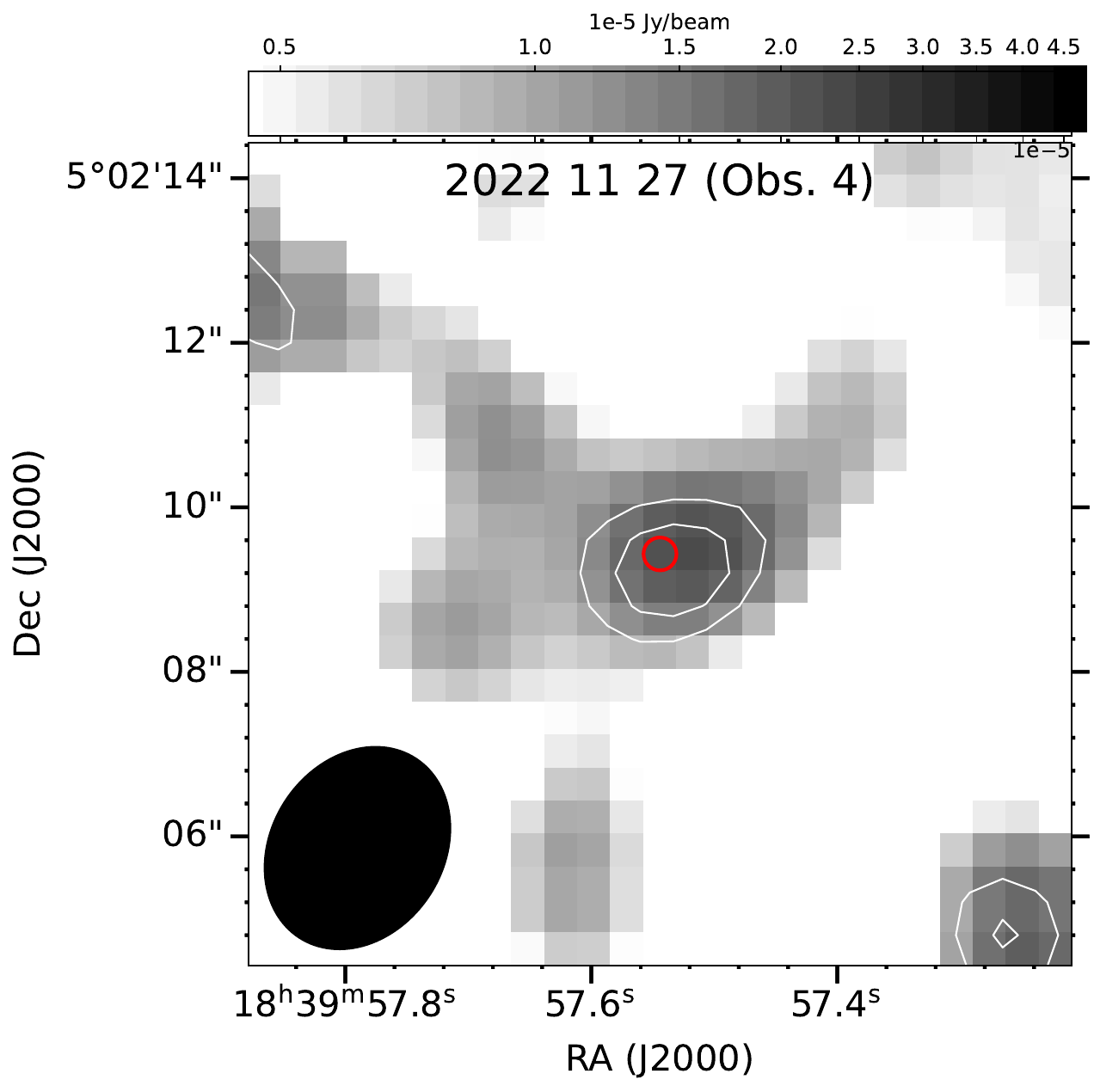}} &
{\includegraphics[width=0.3\textwidth]{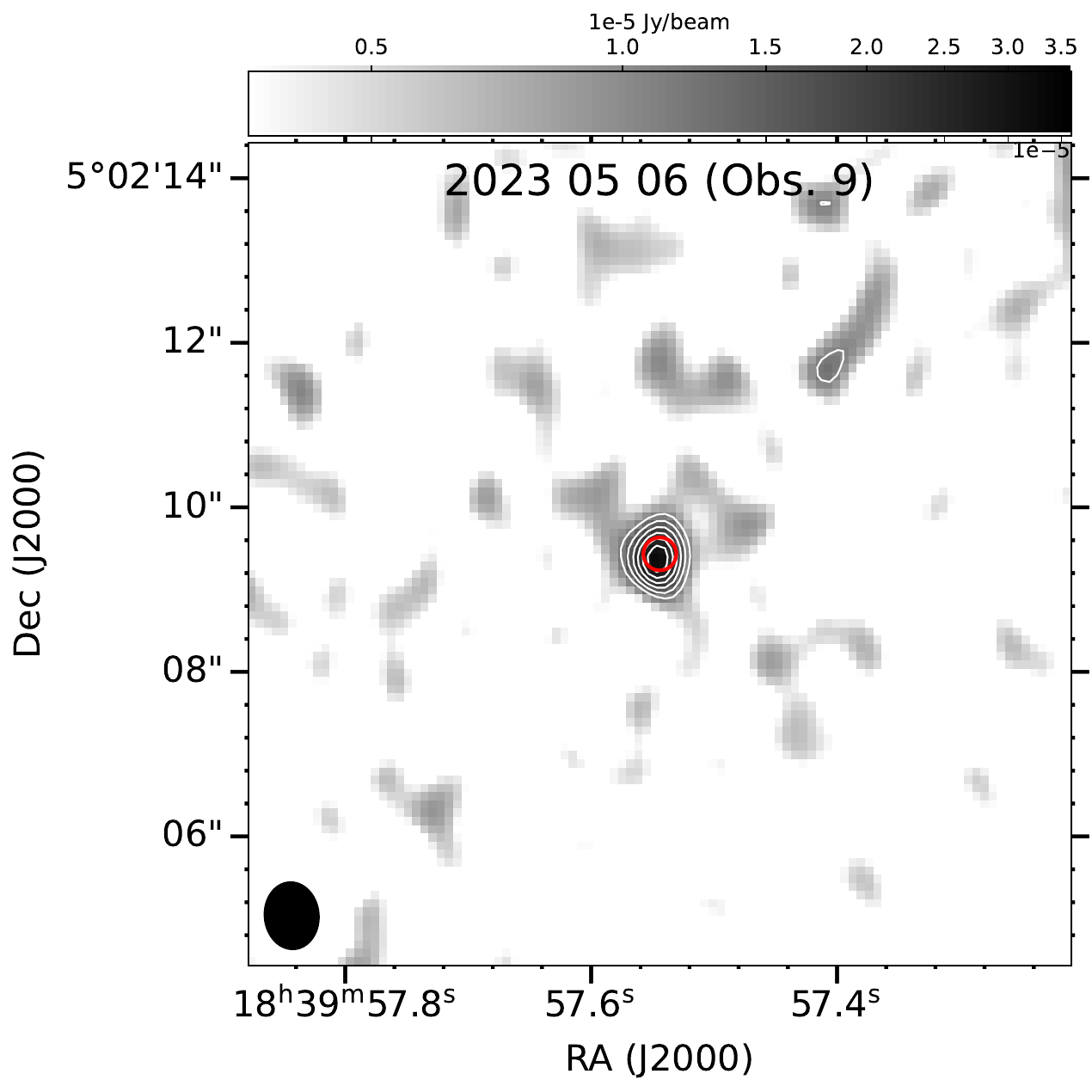}} &
{\includegraphics[width=0.3\textwidth]{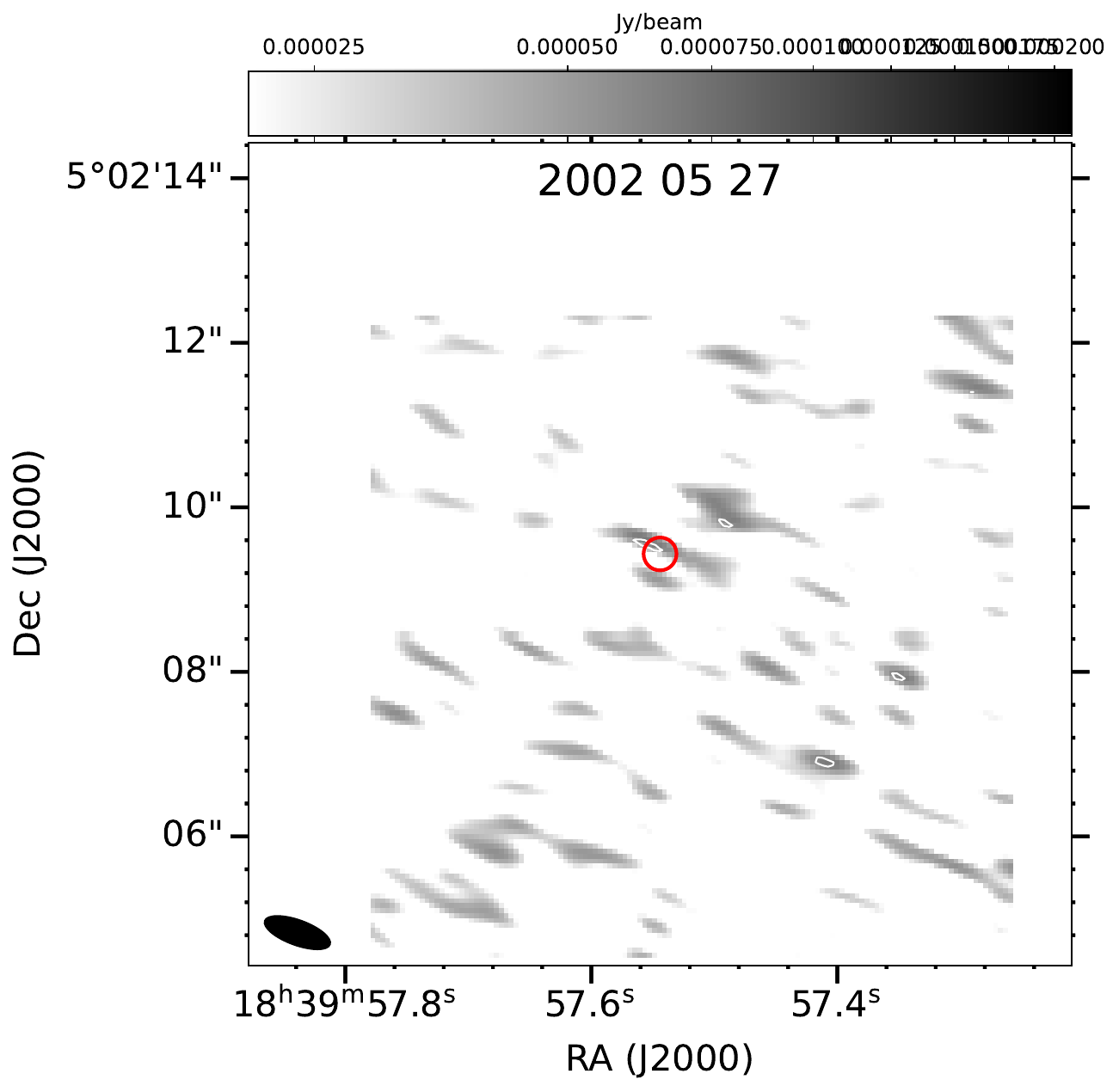}} \\
\end{tabular}
\caption{Images and dates of the radio detections of Ser~X--1: two from new data presented here: {\it left:} Obs. 4; {\it center:} Obs. 9; and {\it right:} the 2002 data presented in \citet{Migliari2004} (data reduced and imaged independently for this manuscript). Contours begin at 3$\sigma$ and increase by 1$\sigma$ (where 1$\sigma$ values are identical to those listed in Table~\ref{table:observations}: (a) 12.6, 16.8 $\mu$Jy; (b) 10.8, 14.4, ... 28.8 $\mu$Jy; (c) 63 $\mu$Jy). The red circle is a 0.2" radius indicating the position of the optical Gaia counterpart to Ser~X--1. The images are identical in angular scale (10" width).}
\label{fig:radio_images}
\end{figure*}

\section{Results and discussion}

Out of 10 total epochs over two years, we have detected Ser~X--1 only twice, in Obs.~4 at $19.9 \pm 4.2~\mu$Jy and in Obs.~9 at $32.2 \pm 3.6~\mu$Jy (Table~\ref{table:observations}; Figure~\ref{fig:MAXIall_inset}). The radio emission is consistent with a point source in each of the detections. Stacking the 8 non-detection observations, we obtain another non-detection with a 3$\sigma$ upper limit of 4.5~$\mu$Jy (i.e. a noise level of 1.5~$\mu$Jy). 

This range of more than a factor of 7 in radio brightness is peculiar given the very consistent X-ray flux level as seen by MAXI, along with the known persistent soft state. Inspecting the MAXI X-ray light curve at each radio epoch (Figure~\ref{fig:MAXIall_inset}) does not offer a pattern that may indicate the reason for intermittent radio emission. Additionally, the hardness-intensity diagram (HID) and color-color diagram (CCD) from MAXI data (Figure~\ref{fig:tracks}), which trace the general evolution of XRB spectral states \citep[e.g.,][]{Hasinger1989, Dunn2010}, also do not indicate that there is any pattern of radio detections within the X-ray data. These diagrams also demonstrate that Ser~X--1 has indeed maintained a very consistent spectral regime over the last decade, specifically a soft state, as there is no extended track evolution present in the daily averaged data.

The radio spectral index (flux density $S_{\nu} \propto \nu^{\alpha}$ where $\alpha$ is the spectral index) of the brighter radio detection is $-1.3 \pm 1.2$ by dividing the 8--12~GHz bandwidth into two subbands. With a large error due to the faintness of the detection, it is challenging to make a clear statement about the spectral index of the emission, but it is consistent with both a flat spectrum compact jet, and steeper spectrum ejecta or other optically thin radio emission. As for polarization, we do not detect either linear or circular polarization in either epoch, resulting in 3$\sigma$ upper limits for polarization degrees of 12.6~$\mu$Jy = $<$63\% and 10.8~$\mu$Jy = $<$34\% for the first and second detections, respectively.

We also investigated the radio emission of the two detections on shorter timescales (Figure~\ref{fig:intra}), to determine if there was strong, short timescale emission present. Inspecting per-scan and 4-scan averages indicates that there is not strong intra-observation variability in either epoch. Particularly, there are no clear detections in any individual scan, ruling out very short, $\sim$minutes timescale burst-like radio emission behavior. The radio emission in each of the two Ser~X--1 epochs appears to indeed occur persistently over the course of the full $\sim$hour, perhaps with some slight drifting, but does not behave as an isolated burst-like event.

Now we discuss possible scenarios that might explain the variable radio emission from Ser~X--1 as it persistently accretes in a soft spectral state.

\subsection{Radio emission due to the companion star}

First we consider scenarios in which the donor star, expected to be an M~dwarf, is directly or indirectly responsible for the observed intermittent radio emission of Ser~X--1.

\subsubsection{Companion star flaring}

A possible interpretation for the intermittent radio emission is that it originates purely from the companion star due to flaring activity. The suggested companion in Ser~X--1 is an M~dwarf, which are well-known hosts of radio flaring. The brightness temperature of the radio emission can be used to test the validity of this possibility. From \citet{Berger2006}: 
\begin{equation}
T_{b} \simeq 2 \times 10^{9} \cdot F_{\nu,\mathrm{mJy}} \cdot \nu^{-2}_{\mathrm{GHz}} \cdot d^{2}_{\mathrm{pc}} \cdot (R/R_{J})^{-2}~\mathrm{K}     
\end{equation}
where $R_{J}$ is the radius of Jupiter, roughly the expected size of the radio emission region in an M~dwarf star. Taking 30~$\mu$Jy flux density at 10~GHz and at a distance of 7.7~kpc, we estimate the brightness temperature from a possible M~dwarf flare to be $\sim 10^{13}$~K. Contrarily, M~dwarf stars have radio brightness temperatures on the order of $10^{8-9}$~K, and are strongly circularly polarized at up to 100\% \citep{Berger2006,Hallinan2008}. Dwarf star flaring also usually occurs over timescales of minutes with an increase by a factor of $\sim$10 or more \citep[e.g.,][]{Berger2001}, rather than relatively stable emission over an hour as we have found from inspecting the data on shorter timescales. 

However, we must also consider that the companion star is bloated and filling its Roche lobe, and is rotating quickly due to tidal locking in the 2 hour orbit with the neutron star. These factors could alter the companion star's radio flaring behavior compared to typical M dwarf stars. \citet{Balona2015} investigated flaring behavior of stars across the Hertzprung-Russell diagram using {\it Kepler} data. Though not radio data, they find that the fraction of flaring stars of a given spectral class increases with rotation rate until a critical saturation value, after which the rate fraction of flaring stars decreases (supersaturation). For M and K dwarf stars, this saturation value was found to be at a 2-3 day rotation period. A rotation period of 2~hours is well past this value and into supersaturation; from these results, it may be unlikely that the companion star of Ser~X--1 flares at all.

\citet{Lim1995} observed radio flares from rapidly rotating G--K dwarf stars in the Pleiades cluster. They found flares lasting a few hours, longer than typical dwarf star flares and more compatible with our results from Ser~X--1. Some flares plateaued at a constant flux density value for $\sim$1~hour. However, these flares are significantly less luminous than the two detections we have for Ser~X--1: the brightest dwarf flare peak luminosity was found to be $2 \times 10^{16}~\mathrm{erg~s^{-1}~Hz^{-1}}$, whereas Ser~X--1 was found to emit at $>10^{18}~\mathrm{erg~s^{-1}~Hz^{-1}}$ in each detection. This discrepancy also casts doubt on the ability for the companion star to be producing the observed radio emission compared to known flaring behaviors of similar but isolated dwarf stars. Thus, though the companion may be of a known flaring class of star, it seems unlikely that the radio emission observed from Ser~X--1 originates in the companion star based on known behaviors of isolated stars or wider binary systems. 

It may still be possible that dwarf star companions in X-ray binaries exhibit unusual flaring activity due to the more extreme environment and rotation rates. If the radio emission in Ser~X--1 is due to the companion star, we can set expectations for seeing a similar behavior in other X-ray binary systems, as there is no reason to believe that Ser~X--1 would be unique in this respect. Given the mass-period relation for main sequence donors in Roche lobe overflow (e.g., \citealt{Knigge2011}), the donor star in Ser~X--1 should be close to 0.2~M$_{\odot}$. 
Thus, if the radio emission is indeed due to a flaring companion, we should expect to see similar intermittent radio emission during times where there is no expected radio jet activity in other short orbital period NSXBs, and even in accreting white dwarf binaries (cataclysmic variables; CVs) where the companion is convective, leading to stellar activity. Similarly, radio emission should not be observed in longer period systems where the companion star is of a more massive spectral class (and hence where there is a radiative envelope), nor in ultracompact systems with white dwarf donors. 

Some radio activity was observed in a few CVs in the Very Large Array Sky Survey (VLASS; \citep{Lacy2020}) \citep{Ridder2023}. These CV radio luminosity levels are a factor of about 5 to 1000 times below that of Ser~X--1, with poor constraints on the duration of the radio emission, and a detection rate of only 0.1\% compared to our 20\% detection rate of Ser~X--1 (for which the probability of detecting 2 out of 10 epochs at an activity rate of 0.1\% is $4\times10^{-5}$). This comparison to CVs, assuming that the radio emission depends purely on the spectral class of companion star that is in a similarly short period accreting binary, also indicates that the radio emission mechanism of Ser~X--1 is distinctly different than that of CVs as a whole, and is not attributable to the donor star itself.


\subsubsection{Shocks with the companion wind}

A final consideration is that the radio emission may arise from a shock between the accretion disk wind and the donor wind. Though a disk wind has not been identified in Ser~X--1 previously, one can indeed be expected due to the persistent soft state nature of the system, and the lack of detection can reasonably be attributed to the system's low inclination. This disk wind is estimated to be $\geq$1000 $\mathrm{km~s^{-1}}$ at a mass loss rate of 10$^{-11}~\mathrm{M_{\odot}}~yr^{-1}$ \citep{Brandt2000,CastroSegura2022,Fijma2023}. M~dwarf wind characteristics vary significantly and are overall poorly constrained; we assume typical mass loss values of 10$^{-13}~\mathrm{M_{\odot}}~yr^{-1}$ and speeds $<1000\mathrm{km~s^{-1}}$ \citep{Wood2001,Wood2021}. From these characteristics, a disk wind would overwhelm the wind from the dwarf companion and not create a shock.

\subsection{Radio emission powered by the accretion process}

Now we consider scenarios in which the radio emission is powered by the neutron star accretor of Ser~X--1.

\subsubsection{Cyclotron maser emission}

A possible radio emission mechanism that has been proposed for a few accreting white dwarf systems is a cyclotron maser\citep{Benz1989,Coppejans2015}, and the face-on inclination of Ser~X--1 may be particularly fortuitous in this case as well. However, cyclotron maser emission is strongly circularly polarized at up to 100\%, and we do not find strong levels of circular polarization present in either of the detections ($<$63\% and $<$34\% for the first and second detections, respectively), thus we believe this is an unlikely explanation for the soft state radio emission.

\subsubsection{Jet precession}

One explanation is that there is indeed a persistent radio jet, but one that is precessing and the resulting change in beaming factor causes the apparent luminosity to vary over long timescales, especially favorable as Ser~X--1 is believed to have a very low inclination. Precessing jets have indeed been observed in a few XRBs \citep{Hjellming1981,Mioduszewski2001,Massi2004}. However, observational data and analysis of NSXBs have indicated that their jets speeds range between $0.3 - 0.5c$ \citep{Fomalont2001,Spencer2013,Coriat2019,Russell2024}, resulting in Lorentz factors of less than $1.2$. This is not a strong enough effect to explain the flux variability factor of $>7$ that has been observed here for Ser~X--1. For this scenario to be plausible, Ser~X--1 would need a highly relativistic jet, far more relativistic than any other NSXB jet known to date, and more comparable to jets of BHXBs. Ser~X--1 behaves very typically as a soft state NSXB, so we do not have a reason to suspect that Ser~X--1 hosts a persistent, highly relativistic and precessing radio jet in its soft state.

\subsubsection{A transient jet}

A more reasonable explanation may be that the observed radio emission is indeed produced by a type of jet during a soft state that is transient. One possibility is that a soft state transient jet is reliant on the boundary layer. Applicable for neutron stars but not black holes, as the accreting material falls toward the neutron star from the inner accretion disk, it may pile up at the equator of the neutron star and form a boundary layer between the inner edge of the accretion disk and the stellar surface \citep{Shakura1988}. The boundary layer can be modelled with X-ray data and past observations of Ser~X--1 have indeed found evidence that its boundary layer can evolve somewhat independently of the mass accretion rate or X-ray luminosity \citep{Chiang2016}. The evolution of the boundary layer cannot be adequately seen in MAXI data, requiring higher sensitivity data to properly model. 
The idea of a jet dependent on the boundary layer has been proposed generally before \citep{Stute2005,Maccarone2008}, and recently invoked specifically and in detail for 4U~1820--30 \citep{Marino2023}, but no detailed theory work has been done in the regime of accretion rates relevant for sources like Ser~X--1. This scenario withstands and would even explain the apparent lack of correlation between the radio detections and MAXI X-ray lightcurve. 
However, without simultaneous sensitive X-ray data during our radio epochs, we cannot yet properly investigate the possibility of a soft state boundary layer and jet coupling in Ser~X--1.

\subsubsection{Relic emission or interstellar medium interaction}

Past observations of radio emission during soft states in X-ray binaries have sometimes been attributed to relic emission from a previous phase of jet activity that has not yet faded away completely (e.g., \citealt{Russell2019}). Since Ser~X--1 has never been observed to be in a hard state, any potential relic emission is not from a canonical hard state compact jet. It may be possible that an intermittent jet of some form discussed above could be produced and then take some time to fade, and this fading stage is what has been observed here with Ser~X--1.

It is also possible that a transient jet is produced, which then interacts with interstellar medium (ISM) surrounding Ser~X--1. This scenario has been invoked for some XRBs in the past to explain persisting radio emission after a jet is no longer expected to be present, or observed deceleration of resolved ejecta \citep[e.g.,][]{Corbel2002,Rushton2017}. Both relic emission and ISM interaction would require an intermittent jet to be produced during the persistent soft state of Ser~X--1, and would also be consistent with a steep spectral index.


\subsection{Radio emission during XRB soft states}

The highly variable or intermittent radio emission of Ser~X--1 presented here also raises the question as to the pervasiveness of soft state radio emission among NSXBs, in both persistently accreting and transient systems. With the relatively low flux densities of a few 10s of $\mu$Jy reported for both Ser~X--1 and 4U~1820--30 during soft states, archival radio data are not sensitive enough to detect similar flux levels in other NSXBs. This motivates new observing campaigns of additional NSXB systems in the future to obtain a better understanding of this sparsely investigated phenomena of soft state radio emission, which BHXBs do not show despite typically being much brighter in radio during other spectral states. It may very well be that neutron stars uniquely are able to display this radio emission phenomenon due to processes involving their solid surfaces and/or intrinsic magnetic fields. 

There is also already an indication of a range of soft state jet behavior among NSXBs. Ser~X--1's emission is intermittent or strongly variable despite very consistent X-ray flux and spectral state as demonstrated in Figures \ref{fig:MAXIall_inset} and \ref{fig:tracks}. 4U~1820--30's radio emission is persistently detected at or above $\sim$30~$\mu$Jy, and overall anti-correlates with its X-ray flux as it slowly evolves in spectral state over a period of a few months \citep{Russell2021}. 4U~1820--30 does also vary in radio luminosity during its soft state by a factor of $\sim$4, but this is accompanied by definitive spectral state and X-ray flux evolution and is thus unsurprising in the context of XRBs overall, in stark contrast to the uncorrelated radio variability we have found here for Ser~X--1. The reason for this difference in radio behavior during soft states between these two persistently active NSXBs is not immediately clear, though it may be related to 4U~1820--30's ultracompact nature or evolution in a globular cluster; a deeper investigation of these two systems may yield more on this topic in the future.

Multiple radio observations of additional systems during soft states with high sensitivity facilities are needed to sample the pervasiveness and general characteristics of NSXB soft state radio emission. If soft state radio emission is indeed attributed to the boundary layer, simultaneous X-ray observations would also be needed to model the boundary layer simultaneously with the radio emission to further investigate the coupling between them. This would also be another phenomenon of NSXBs specifically that expands the wide range of behaviors that these accreting neutron stars are known to exhibit and motivate additional theoretical work on accretion regimes at moderately high accretion rates.

\section{Summary}

We have detected radio emission from the persistent soft state NSXB Ser~X--1 in 2 out of 10 total epochs over two years. Ser~X--1's X-ray flux and spectral state has been very constant over this time period as observed by MAXI, and we find no pattern between the radio detections and X-ray properties. The non-detections' stacked limit indicates that Ser~X--1 is capable of varying in radio by a factor of $>$7 from our recent data, or by even more including the $\sim$80~$\mu$Jy detection by \citet{Migliari2004} \footnote{This detection is at a bit less than 4$\sigma$, and is located on a local positive noise region that is not a well-shaped point source, so caution is warranted in interpreting it; see Figure~\ref{fig:radio_images}.}, while the X-ray flux only varies by a factor of about 2. We have considered several explanations for our results, and excluded the possibility that the radio emission originates from a flaring donor star, preferring that the variable radio emission originates from a transient jet. Soft state radio emission from NSXBs is poorly investigated and similarly not well understood, as the only other system to date with multiple soft state radio observations and detections is 4U~1820--30. As no BHXBs have definitively shown active soft state radio emission, the NSXB mechanism is likely related to the NS itself, such as its solid surface or magnetic field. Tentative theories have indeed attributed NS soft state radio emission to activity in the boundary layer of the NS. We believe that the boundary layer could also be the driving factor here in Ser~X--1, as its evolution has been observed to occur independently of the X-ray flux. A future simultaneous X-ray and radio campaign will be able to investigate this possibility by thoroughly modelling its boundary layer properties from X-ray data during active and inactive radio epochs.




\begin{acknowledgments}
We thank the anonymous referee for thoroughly reading the manuscript and providing helpful feedback. We thank Tracy Clarke for providing information on the commensal VLITE data for our observations, and Navin Sridhar for discussions of neutron star boundary layers. AJT acknowledges the support of the Natural Sciences and Engineering Research Council of Canada (NSERC; funding reference number RGPIN--2024--04458). The National Radio Astronomy Observatory is a facility of the National Science Foundation operated under cooperative agreement by Associated Universities, Inc.

\end{acknowledgments}

%

\vspace{5mm}
\facilities{Very Large Array (VLA)}


\software{Common Astronomy Software Applications (CASA, \citet{CASA2022}); Astronomical Plotting Library in Python (APLpy, \citet{APLPY2012})
          }



\appendix

\section{Positions and \texttt{uvmodelfit}}
We report the position of the peaks of the two radio detections. \texttt{imfit} did not perform well to recover the position of the two detections, with position errors of $>1$" for synthesized beam sizes of $\lesssim2$". We also used \texttt{uvmodelfit} as a second method of obtaining the flux densities of the two detections, and the results of this method indeed agree with the flux density values obtained from standard imaging methods. 

\begin{table*}[h!]
\centering
    \setlength{\tabcolsep}{10pt}
	\begin{tabular}{cccc}
	    \hline
	    Obs. \# & RA & Dec & \texttt{uvmodelfit}  \\
               &    &     & I ($\mu$Jy) \\
		\hline\hline
        4 & 18:39:57.545 & +05:02:09.164 & $24.1\pm2.4$  \\   
        9 & 18:39:57.547 & +05:02:09.391 & $33.1\pm2.1$  \\
        \hline	
	\end{tabular}
	\caption{Positions and \texttt{uvmodelfit} flux densities (Stokes I) of the two Ser~X--1 detections. Positions were obtained from the peak location in an over-sampled cellsize image (\texttt{imfit} did not perform well to obtain positions of the two detections). \texttt{uvmodelfit} fit a point source in 5 iterations to the visibility data after using \texttt{uvsub} to remove other sources in the field of view. The \texttt{uvmodelfit} flux densities are indeed consistent within $1\sigma$ with the values obtained from imaging.}
	\label{table:fits}
\end{table*}

\section{Other radio sources in the field of view}
We report the presence of 5 additional detected radio sources in the field of view in the stacked data at 10~GHz in Table~\ref{table:otherSources}, all of which are consistent with being point sources at the highest observed angular resolution of $\sim$ 0.6". 

\begin{table*}[h!]
\centering
    \setlength{\tabcolsep}{10pt}
	\begin{tabular}{ccccc}
	    \hline
	    J2000 & RA & Dec & Peak mJy/beam & Local RMS $\mu$Jy/beam\\
		\hline\hline
		J1840+0505 & 18:40:03.078 & +05:05:00.892 & 4.987 & 15.5 \\
        J1839+0459 & 18:39:52.054 & +04:59:48.793 & 0.123 & 5.4 \\
        J1839+0500 & 18:39:52.527 & +05:00:51.391 & 0.030 & 2.6  \\
        J1840+0502 & 18:40:10.700 & +05:02:47.921 & 0.107 & 10.3 \\
        J1839+0503 & 18:39:54.376 & +05:03:03.648 & 0.028 & 1.9 \\   
        \hline	
	\end{tabular}
	\caption{Table of sources other than Ser~X--1 in the field of view of the observations. Values here are obtained from the primary beam corrected image of stacking Ser~X--1 non-detection data.}
	\label{table:otherSources}
\end{table*}


\bibliography{example}{}
\bibliographystyle{aasjournal}



\end{document}